\documentclass[runningheads]{llncs}
\usepackage[T1]{fontenc}
\usepackage{graphicx}
\usepackage{booktabs}
\usepackage{xcolor}
\usepackage{colortbl}
\usepackage{amsmath,amssymb,amsfonts}
\usepackage{multirow}
\usepackage{marvosym}
\usepackage[colorlinks,linkcolor=red]{hyperref}

\begin{document}
%
% \title{A new dataset and a baseline model for Video Colorectal Cancer Segmentation in Endorectal Ultrasound}
\title{Towards a Benchmark for Colorectal Cancer Segmentation in Endorectal Ultrasound Videos: Dataset and Model Development}
\titlerunning{Towards a Benchmark for CRC Segmentation in ERUS Videos}
% If the paper title is too long for the running head, you can set
% an abbreviated paper title here
%
\author{
    Yuncheng Jiang\inst{1,2,3,4\footnotemark[4]}  \and   % index{Jiang, Yuncheng}
    Yiwen Hu\inst{1,2\footnotemark[4]} \and     % index{Hu, Yiwen}
    Zixun Zhang\inst{1,2,3\footnotemark[4]} \and   % index{Zhang, Zixun}
    Jun Wei\inst{1,2} \and   % index{Wei, Jun}
    \\
    Chun-Mei Feng\inst{3} \and % index{Feng, Chunmei}
    Xuemei Tang\inst{5} \and % index{Tang, Xuemei}
    Xiang Wan\inst{4} \and   % index{Wan, Xiang}
    \\
    Yong Liu\inst{3} \and   %index{Liu Yong}
    Shuguang Cui\inst{1,2} \and  % index{Cui, Shuguang}
    Zhen Li\inst{1,2}\textsuperscript{\Letter}  % index{Li, Zhen}
}
\authorrunning{Y. Jiang et al.}
% First names are abbreviated in the running head.
% If there are more than two authors, 'et al.' is used.
%
\institute{
SSE, The Chinese University of Hong Kong, Shenzhen, China
\and
Shenzhen Future Network of Intelligence Insititute, China
\and
Institute of High Performance Computing, Agency for Science, Technology and Research, Singapore
\and
Shenzhen Research Institute of Big Data, Shenzhen, China
\and
Affiliated Hospital of North Sichuan Medical College, Sichuan, China\\
\email{yunchengjiang@link.cuhk.edu.cn, lizhen@cuhk.edu.cn}}
\maketitle              % typeset the header of the contribution
\renewcommand{\thefootnote}{\fnsymbol{footnote}}
\footnotetext[4]{Equal contributions.}
% \footnotetext[1]{Corresponding authors.}

\begin{abstract}

Endorectal ultrasound (ERUS) is an important imaging modality that provides high reliability for diagnosing the depth and boundary of invasion in colorectal cancer. However, the lack of a large-scale ERUS dataset with high-quality annotations hinders the development of automatic ultrasound diagnostics. In this paper, we collected and annotated the first benchmark dataset that covers diverse ERUS scenarios, \textit{i.e.} colorectal cancer segmentation, detection, and infiltration depth staging. Our ERUS-10K dataset comprises 77 videos and 10,000 high-resolution annotated frames. Based on this dataset, we further introduce a benchmark model for colorectal cancer segmentation, named the \textbf{A}daptive \textbf{S}parse-context \textbf{TR}ansformer (\textbf{ASTR}). ASTR is designed based on three considerations: scanning mode discrepancy, temporal information, and low computational complexity. For generalizing to different scanning modes, the adaptive scanning-mode augmentation is proposed to convert between raw sector images and linear scan ones. For mining temporal information, the sparse-context transformer is incorporated to integrate inter-frame local and global features. For reducing computational complexity, the sparse-context block is introduced to extract contextual features from auxiliary frames. Finally, on the benchmark dataset, the proposed ASTR model achieves a $77.6\%$ Dice score in rectal cancer segmentation, largely outperforming previous state-of-the-art methods. 
% Codes are available at:~\href{https://github.com/yuncheng97/ASTR.git}{https://github.com/yuncheng97/ASTR.git}

\end{abstract}

\section{Introduction}
\label{sec:introduction}
\begin{figure}[t]
    \centering
    \includegraphics[width=0.8\linewidth]{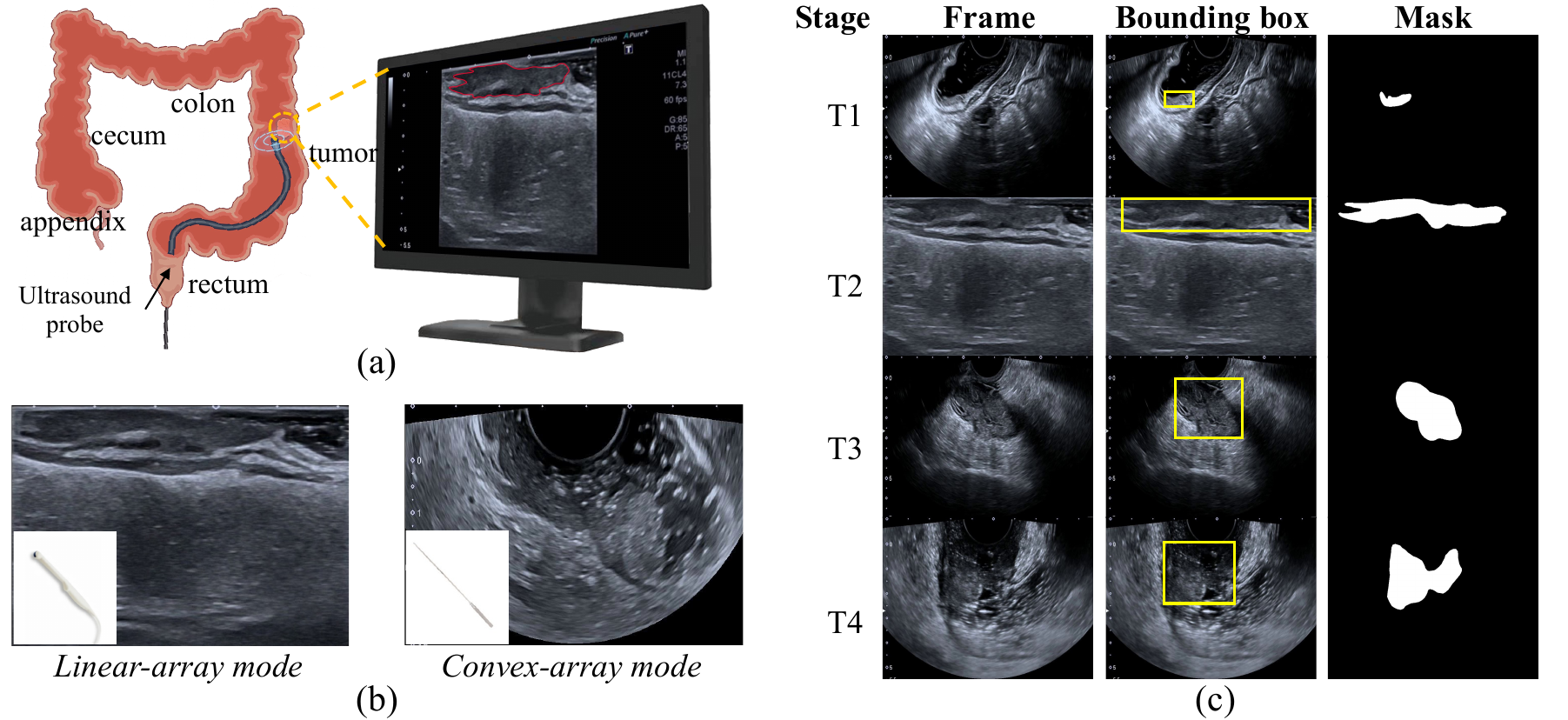}
    \caption{(a) Schematic diagram of ERUS operation. (b) Different scanning modes of ultrasound. (c) Examples of our ultrasound video dataset with corresponding labels.}
    \label{fig:teaser}
\end{figure}

Colorectal cancer (CRC) has become the second leading cause of cancer death worldwide~\cite{favoriti2016worldwide}.
Accurate early detection of CRC is crucial for making therapeutic decisions and improving the survival rate.
Currently, Endorectal ultrasound (ERUS) is adopted as a routine imaging modality for diagnosing and staging colorectal cancer~\cite{hunerbein2003endorectal}.
As shown in Fig.~\ref{fig:teaser}(a), ERUS provides in-depth assessments of tumor infiltration, precisely depicting the cancer's location, size, and its relationship with surrounding tissues~\cite{rieger2004endoanal}.
However, the sonographers's level of experience or fatigue during long duty hours contributes to a non-negligible rate of missed detections in clinical diagnosis.
Thus, it is important to develop an automatic system for computer-aided diagnosis of CRC from ERUS videos. 

%%%%%%%%%%%%%%%%%%%%%%%%%%% Fig 1 %%%%%%%%%%%%%%%%%%%%%%%%%
%%%%%%%%%%%%%%%%%%%%%%%%%%% Fig 1 %%%%%%%%%%%%%%%%%%%%%%%%%

Previous works have proposed well-annotated ultrasound datasets and corresponding methods, covering various organs such as the breast, pancreas, and thyroid~\cite{han2023thyroid,li2023dsmt,lin2023shifting}.
Li \textit{et al.}~\cite{li2023dsmt} pioneer the application of deep neural networks in the diagnosis of pancreatic diseases using Endoscopic Ultrasound (EUS). 
Wang \textit{et al.} introduce a 3D feature pyramid network to conduct inter-frame collaboration.
Li \textit{et al.}~\cite{li2022rethinking} propose a memory bank and dynamically update the memory to establish long-term temporal correlation.
Building upon this, Lin \textit{et al.}~\cite{lin2023shifting} extend the approach by incorporating Fourier transforms to aggregate multiple features from the frequency domain.
Despite significant advancements achieved by those methods, accurate colorectal cancer segmentation in ERUS remains challenging due to 
(1) the scarcity of large-scale endorectal ultrasound datasets to train well-converged segmentation models, 
(2) the intrinsic scanning mode discrepancy obtained from different ultrasound sensors, as illustrated in Fig.~\ref{fig:teaser}(b),
and (3) the motion blur resulted from a rapidly moving ultrasound probe.

To address the above issues, we collected 77 endorectal ultrasound videos with 10,000 well-annotated frames and propose the \textbf{\textit{first} large-scale ERUS video colorectal cancer segmentation dataset,} of which 57 videos contain tumor infiltration depth staging, contributing to realistic clinical scenarios.
Apart from the benchmark dataset, we further propose a benchmark model for colorectal cancer segmentation, termed the \textbf{A}daptive \textbf{S}parse-context \textbf{TR}ansformer (\textbf{ASTR}).
ASTR is designed based on three main considerations: scanning mode discrepancy, temporal information, and computational complexity. 
To generalize to different scanning modes, the adaptive scanning mode augmentation (ASMA) is proposed, which converts images between linear and convex scanning modes through coordinate transformation ({\it i.e.}, Polar and Cartesian coordinate systems).
To exploit temporal information between frames, the Sparse-context Transformer is incorporated to integrate inter-frame local and global features. 
To reduce computational complexity during temporal fusion, the
Sparse-context Block (SCB) is introduced to extract contextual features while eliminating irrelevant noises from reference frames.

In summary, our contributions include three aspects: 
(1) We present the first well-annotated endorectal ultrasound dataset with comprehensive annotations, laying the foundation for advancements in automatic ultrasound diagnosis of colorectal cancer. 
(2) We build the first adaptive sparse-context transformer model tailored for video colorectal cancer segmentation. 
(3) We have conducted extensive experiments and established a comprehensive benchmark for video colorectal cancer segmentation. Experimental results demonstrate that our proposed method achieves state-of-the-art performance. 

\section{Method}
\label{sec:method}

\subsection{Adaptive Scanning Mode Augmentation}
In practical endorectal ultrasound examinations, sonographers employ different probes depending on specific patient conditions. These probes exhibit varying scanning modes ({\it i.e.}, linear and convex) and result in images of different forms. To ensure our model can generalize to these images under different modes,
we propose a simple yet effective data augmentation method called Adaptive Scanning Mode Augmentation (ASMA). 

For the convex-array and linear-array modes, there exists a coordinate mapping relationship between them. 
Therefore, we can obtain the image transformation between different modes through Polar-Cartesian transformation.
Specifically, we establish the polar coordinates for the convex-array mode image with the origin point $(r_o, \theta_o)$ set at the top center of the image and the x-axis at the top edge.
Meanwhile, Cartesian coordinates are set with $(x_o, y_o)$ as the origin, the top edge as the x-axis, and the vertical central axis as the y-axis. 
\textbf{For the convex-array mode}, we obtain the value and position of each pixel $(x,y)$ on the transformed image according to its corresponding pixel $(r,\theta)$ on the original convex-array mode image by the following polar-to-cartesian transformation:
\begin{equation}
    \begin{aligned}
    x= x_o+r\cos\theta, \quad y=y_o+r\sin\theta.
    \end{aligned}   
    \label{equ:1}
\end{equation}

\noindent Similarly, \textbf{for the linear-array mode}, the transformed position $(r,\theta)$ of convex-array mode image is calculated using the cartesian-to-polar transformation:
\begin{equation}
    \begin{aligned}
    r= \sqrt{x^2+y^2}, \quad \theta=\arctan(\frac{y}{x}).
    \end{aligned}   
    \label{equ:2}
\end{equation}
 
After augmentation, the number of images in the original dataset will be balanced across different scanning modes, reducing the risk of model overfitting to any particular mode due to the imbalanced dataset.

%%%%%%%%%%%%%%%%%%%%%%%%%%% Fig 3 %%%%%%%%%%%%%%%%%%%%%%%%%
\begin{figure}[t]
    \centering
    \includegraphics[width=0.9\linewidth]{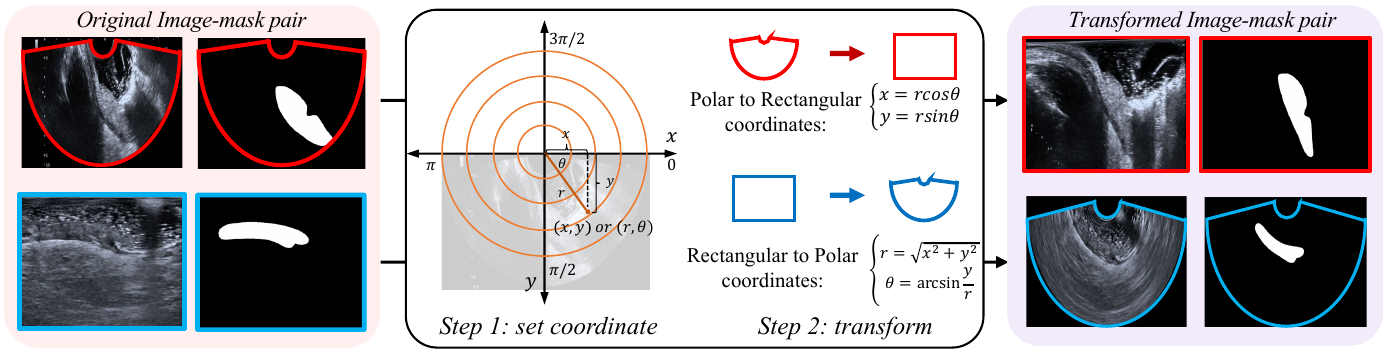}
    \caption{Schematic illustration of the adaptive scanning mode augmentation (ASMA). The original frame of linear-array/convex-array mode is transformed to the frame of convex-array/linear-array mode by Polar-Cartesian coordinate system transformation, enhancing the model's generalization ability on different scanning modes.}
    \label{fig:module}
\end{figure}
%%%%%%%%%%%%%%%%%%%%%%%%%%% Fig 3 %%%%%%%%%%%%%%%%%%%%%%%%%

\subsection{Sparse-context Transformer}

\textbf{Per-frame context learning.}
The Sparse-context Transformer can be divided into two key stages: context learning and context fusion. 
In the per-frame context learning phase, coarse per-frame context feature $f_t\in\mathbb
{R}^{c\times h\times w}$ are initially extracted through the backbone network. 
Then, we refine $f_t$ via stacked transformer layers with self-attention (SA) operation:
\begin{equation}
    \begin{aligned}
        SA(f_t)= \operatorname{softmax}(\frac{q(f_t)k(f_t)^T}{\sqrt{d}})v(f_t), \quad  p=SA(f_t),
        % MSA(T)    &=Concat(SA_1,SA_2,\cdots,SA_h)W^H
    \end{aligned}
    \label{equ:2}
\end{equation}
where $p\in \mathbb{R}^{c\times h \times w}$ is the refined per-frame context and $q,k,v$ are the linear projection functions, \textit{i.e.} $q(f_t) \in \mathbb{R}^{c \times d}$.

\noindent \textbf{Reference-frame context learning.}
For the reference-frame context learning stage, we enhance the coarse backbone features 
$\{f_{t-1},f_{t-2},\cdots,f_{t-T}\}$ via the sparse-context block (SCB), producing sparse representations for reference-frame contexts $r_{t-1}\in\mathbb{R}^{c\times n_{t-1}}, \cdots, r_{t-T} \in\mathbb{R}^{c\times n_{t-T}} $. 
Those features are concatenated along the channel dimension to form the reference-frame context $r \in\mathbb{R}^{c\times m}, ~m=(n_{t-1}+\cdots+n_{t-T})$.
Then, in the context fusion stage, we adopt a temporal transformer architecture equipped with cross-attention (CA) as described in Li~\textit{et al.}\cite{li2021video} to generate multi-frame contexts $y$:
\begin{equation}
    \begin{aligned}
        CA(p, r)= \operatorname{softmax}(\frac{q(p)k(r)^T}{\sqrt{d}})v(r), \quad y= CA(p, r).
        % MSA(T)    &=Concat(SA_1,SA_2,\cdots,SA_h)W^H
    \end{aligned}
    \label{equ:2}
\end{equation}
Finally, the fused features are fed to a sequential of up-sampling layers and a segmentation head to generate the predicted mask.

%%%%%%%%%%%%%%%%%%%%%%%%%%% Fig 2 %%%%%%%%%%%%%%%%%%%%%%%%%
\begin{figure}[t]
    \centering
    \includegraphics[width=\linewidth]{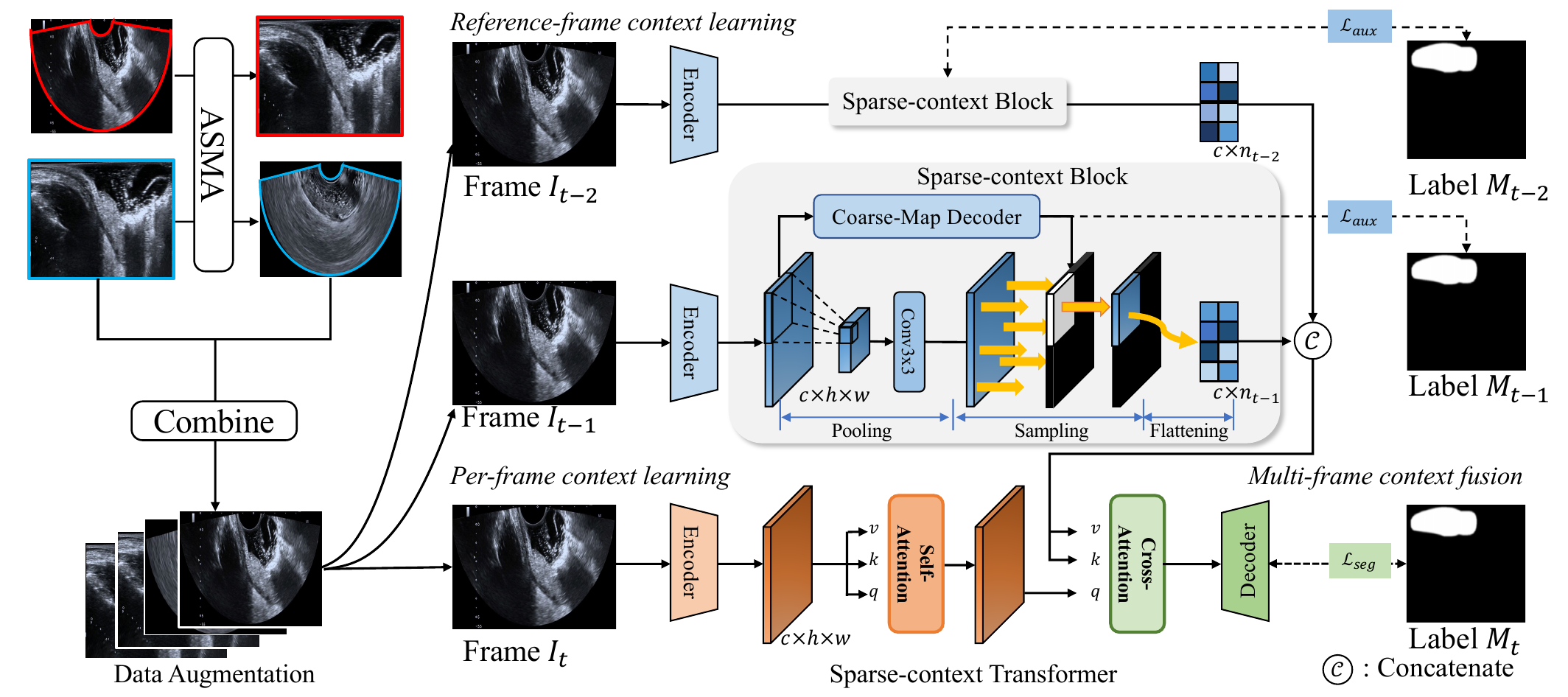}
    \caption{Pipeline of the proposed ASTR. To generalize to different scanning modes, we first conduct data augmentation by interconverting the linear-array mode and convex-array mode in the adaptive scanning mode augmentation (ASMA). During training, the Sparse-context Transformer extracts inter-frame contexts to exploit spatiotemporal information. Furthermore, we devise a Sparse-context Block to eliminate the irrelevant background noise and reduce computational cost. Finally, the multi-frame contexts from all samples are fused for segmentation mask prediction.}
    \label{fig:framework}
\end{figure}
%%%%%%%%%%%%%%%%%%%%%%%%%%% Fig 2 %%%%%%%%%%%%%%%%%%%%%%%%%

\noindent \textbf{Sparse-context Block.}
Historical evidence suggests that adjacent frames encapsulate valuable temporal contexts, thereby aiding in bolstering the model's generalization capabilities.
However, these temporal contexts are sparsely distributed in adjacent frames~\cite{sun2022coarse}. Namely, most pixels are redundant, thereby increasing the computational burden unnecessarily.
To reduce the computational cost, we summarized two sparse principles:
(1) nearby reference frames provide limited distinguished information,
(2) target regions are crucial information for learning. 
Based on these two principles, we have devised a sparse-context block for selecting key information from reference features. 
Following principle (1), features undergo pooling operations to adjust the receptive field, followed by a $3\times 3$ convolutional layer to refine the features. Here, the pooling size $K$ is determined by proximity: $K=2^i$, where i is the index of the reference frame. 
This process ensures that distant frames can learn features from a larger receptive field, contributing meaningful information. 
Following principle (2), we introduce explicit attention to obtain sparse representations. 
Specifically, features pass through a coarse-map decoder, which is a $1\times 1$ convolution layer, to generate a coarse mask of the lesion region. 
We sample feature values at positions corresponding to the non-zero position in the coarse mask, ensuring that only regions likely to contain lesions contribute to subsequent multi-frame context fusion. 
Through sparse representation, the computational complexity in the context fusion stage decreases from $\mathcal{O}(h^2w^2tc)$ to $\mathcal{O}(hwmc)$, where $m \ll hwt$. 

\subsection{Loss Function}
We adopt the combination of binary cross-entropy loss $\mathcal{L}_{bce}$, Dice loss $\mathcal{L}_{dice}$, and mean absolute error loss $\mathcal{L}_{mae}$ for pixel-level supervision. 
Additionally, to refine the localization of lesion regions within the sparse attention module, we incorporate an auxiliary loss function $\mathcal{L}_{aux}$ to supervise the coarse-map decoder.
This auxiliary loss, using the ground truth label as supervision, ensures more accurate guidance for SCB during the feature selection process, aligning the model's focus with the actual lesion areas.
The total loss function is defined as follows:
\begin{equation}
    \begin{aligned}
    \mathcal{L}_{seg} = \mathcal{L}_{aux}&= \mathcal{L}_{bce} + \mathcal{L}_{dice} + \mathcal{L}_{mae},\\
    \mathcal{L}_{total} &=\mathcal{L}_{seg} + \lambda_{aux} \cdot \mathcal{L}_{aux},
    \end{aligned}
\end{equation}
where the hyper-parameter $\mathcal{L}_{aux}=0.3$ is used to balance the weight between the segmentation loss and auxiliary loss.

%%%%%%%%%%%%%%%%%%%%%%%%%%% Fig 2 %%%%%%%%%%%%%%%%%%%%%%%%%
\begin{figure}[t]
    \centering
    \includegraphics[width=\linewidth]{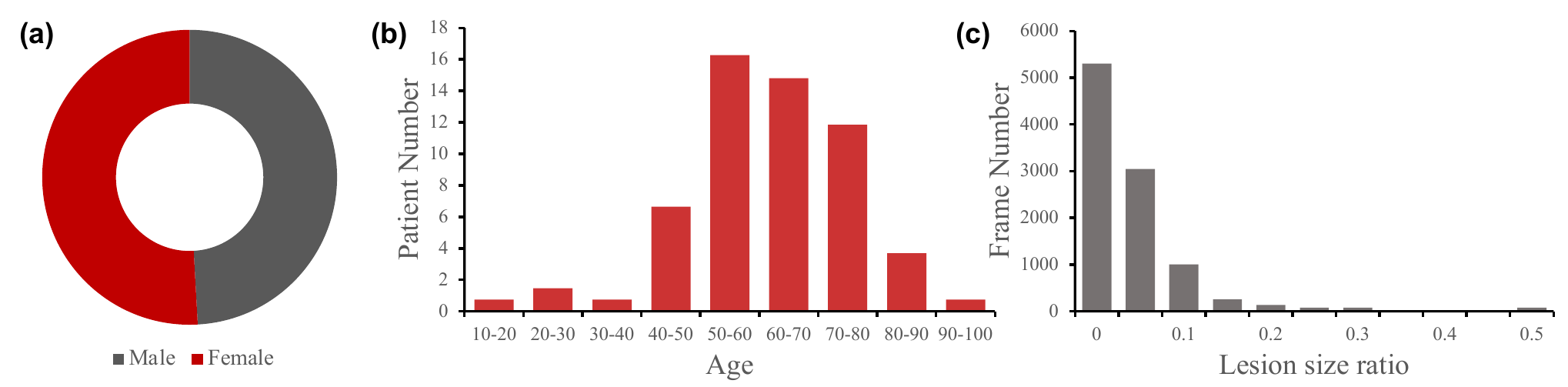}
    \caption{Dataset statistic. (a) Gender distribution. (b) Age distribution. (c) Lesion size distribution.}
    \label{fig:data_stat}
\end{figure}
%%%%%%%%%%%%%%%%%%%%%%%%%%% Fig 2 %%%%%%%%%%%%%%%%%%%%%%%%%

\section{Experiment}
\label{sec:experiment}

\textbf{Endorectal Ultrasound Dataset}
To facilitate advancements in colorectal cancer segmentation and the staging of tumor infiltration depth, we collected and annotated the first endorectal ultrasound dataset, named \textbf{ERUS-10K}, consisting of 77 endorectal ultrasound videos with a total of 10,000 annotated frames.
All patients underwent endorectal ultrasound examinations at the Affiliated Hospital of North Sichuan Medical College using CANNO-type color Doppler ultrasound diagnostic apparatus. 
Among 77 videos, 19 videos were recorded using the linear-array scanning mode by 11CL4 rectal cavity probe, while the remaining utilized the convex-array scanning mode by the vaginal probe. 
Fig.~\ref{fig:teaser}(b) illustrates the differences between these two scanning modes. 
Manual annotations of colorectal cancers were performed by experienced sonographers.
The provided annotations include colorectal lesion masks and bounding boxes, comprehensively covering clinical scenarios ranging from colorectal lesion detection to segmentation.
Furthermore, 57 videos implemented pathological examinations via percutaneous biopsy to determine the tumor infiltration depth (\textit{i.e.} stage T1, T2, T3, T4), laying the foundation for automated and precise colorectal cancer staging. 
Fig.~\ref{fig:teaser}(c) displays sample images along with their corresponding labels. 
The entire dataset is divided into training, validation, and test sets in a ratio of 7:1:2, enabling a comprehensive benchmark evaluation of our proposed methods.
Fig.~\ref{fig:data_stat} shows the detailed statistics of the dataset.

\noindent \textbf{Implementation Details.}
Our network was implemented using the PyTorch framework on two NVIDIA V100 GPUs. We adopt the Res2Net-50~\cite{gao2019res2net} pre-trained on ImageNet as the backbone. 
We sample a video clip with $T=3$ frames for training and inference.
All input images are uniformly resized to $352\times352$ and employ random flip as data augmentation. 
The whole network is trained in an end-to-end manner using Adam optimizer.
The initial learning rate is set to 0.0001.
We train the entire model for 24 epochs with batch size 24.

\subsection{Comparisons with State-of-the-arts}
We conduct comparisons between our method with \textbf{10} state-of-the-art segmentation methods, comprising \textbf{5} image-based methods: UNet~\cite{ronneberger2015u}, SANet~\cite{wei2021shallow}, TransUNet~\cite{chen2021transunet}, SETR~\cite{zheng2021rethinking}, and MedSAM~\cite{ma2024segment}, and \textbf{5} video-based methods: DAF-Net~\cite{wang2019deep}, PNS-Net~\cite{ji2021progressively}, DCF-Net~\cite{zhang2021dynamic}, SLT-Net~\cite{cheng2022implicit}, and FLA-Net~\cite{lin2023shifting}.
We compare the performances using various metrics, including the Dice similarity (Dice), Intersection over Union (IoU), Mean Absolute Error (MAE), Sensitivity (Sen), Specificity (Spe), and inference Frame Per Second (FPS).

\noindent \textbf{Quantitative Comparisons.} As shown in Table 1, among the compared methods, video-based methods generally outperformed image-based methods since temporal information is considered. 
With the proposed modules, our ASTR model exhibited significant improvements across all metrics, surpassing the state-of-the-art methods by a large margin.
Specifically, it increased the Dice score from $75.7\%$ to $77.6\%$, the IoU score from $62.8\%$ to $65.7\%$, the sensitivity from $76.0\%$ to $78.5\%$, and reduced the MAE score from $2.9\%$ to $2.7\%$.

%%%%%%%%%%% Tab: quantitative results %%%%%%%%%%%%%%%

\begin{table*}[t]
    \centering
    \caption{Quantitative comparison of our ASTR and other state-of-the-art methods on the ERUS video lesion segmentation dataset.}
    \renewcommand\arraystretch{1}
    \begin{tabular}{crlcccccc}
    \toprule[1pt]
    \makebox[0.05\textwidth][c]{} & \makebox[0.1\textwidth][c]{Pubs.} & \makebox[0.1\textwidth][c]{Methods} & \makebox[0.09\textwidth][c]{MAE$\downarrow$} & \makebox[0.09\textwidth][c]{IoU$\uparrow$}  & \makebox[0.09\textwidth][c]{Dice$\uparrow$} & \makebox[0.09\textwidth][c]{Sen$\uparrow$}  & \makebox[0.09\textwidth][c]{Spe$\uparrow$}  & \makebox[0.09\textwidth][c]{FPS$\uparrow$} \\
    \midrule
    % \textit{MICCAI15}  & UNet\cite{ronneberger2015u}    
    \multirow{5}*{\rotatebox{90}{Image-based}} 
    & \textit{MICCAI15}  & UNet
     & 5.0 & 59.3 & 72.6 & 74.4 & 98.1 & 65.2       \\
    % \textit{MICCAI20}  & SANet\cite{wei2021shallow}   
    &\textit{MICCAI20}  & SANet 
     & 4.5 & 59.8 & 73.2 & 72.0 & 98.5 & 56.6      \\
    % \textit{CVPR21}    & TransUNet\cite{chen2021transunet} 
    &\textit{CVPR21}    & TransUNet
     & 5.2 & 58.0 & 71.7 & 70.9 & 98.0 & 58.1     \\
    % \textit{CVPR21}    & SETR\cite{zheng2021rethinking}   
    &\textit{CVPR21}    & SETR
     & 7.4 & 56.0 & 69.8 & 75.1 & 97.8 & 27.5    \\
    % \textit{MIA2024}    & SAM\cite{huang2024segment} 
    &\textit{Nature24}    & MedSAM
     & 2.9 & 62.8 & 75.7 & 75.5& 98.3 & 2.2    \\ \midrule
    % \textit{TMI19}    & DAF-Net\cite{wang2019deep}    
    \multirow{5}*{\rotatebox{90}{Video-based}} & \textit{TMI19}    & DAF-Net
     & 4.2 & 59.9 & 73.6 & 73.7 & 98.1 & 30.1     \\
    % \textit{MICCAI21}  & PNS-Net\cite{ji2021progressively} 
    & \textit{MICCAI21}  & PNS-Net
     & 3.6 & 61.2 & 74.5 & 75.8 & 98.3 & 16.5     \\
    % \textit{CVPR21}    & DCFNet\cite{zhang2021dynamic}  
    & \textit{CVPR21}    & DCF-Net
     & 3.2 & 61.6 & 74.7 & 73.2 & 98.4 & 6.6    \\
    % \textit{CVPR22}    & SLT-Net\cite{cheng2022implicit} 
    & \textit{CVPR22}    & SLT-Net
     & 3.0 & 62.2 & 75.2 & 74.0 & 98.4 & 10.2     \\
    % \textit{MICCAI23}  & FLA-Net\cite{lin2023shifting} 
    & \textit{MICCAI23}  & FLA-Net
     & 3.2 & 61.7 & 74.8 & 76.0 & 98.3 & 35.5     \\ \midrule
    \rowcolor{gray!20} 
    && \textbf{ASTR}        & \textbf{2.7} & \textbf{65.7} & \textbf{77.6} & \textbf{78.5} & \textbf{98.6} & \textbf{40.8}\\
    \bottomrule[1pt]
    \end{tabular}
    \label{tab:polyp_results}
\end{table*}

\begin{figure*}[t]
    \centering
    \includegraphics[width=0.9\linewidth]{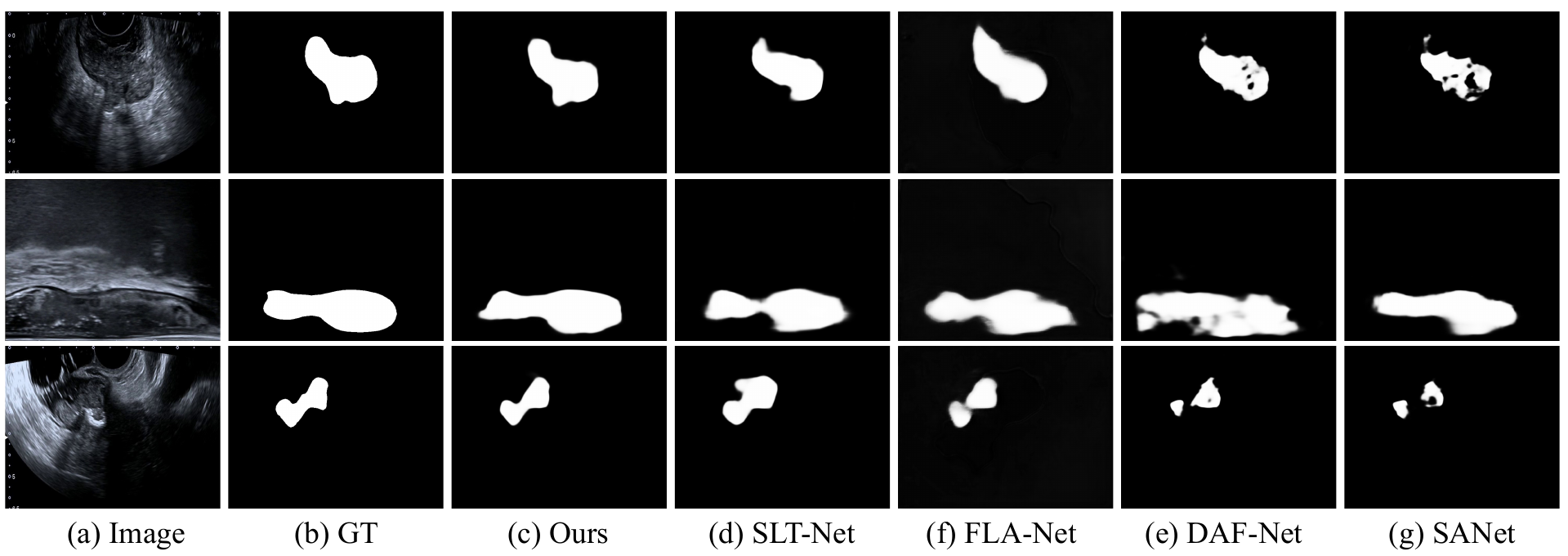}
    \caption{\textbf{Qualitative comparisons}. "GT" denotes the ground truth. See \textit{Suppl.} for more visualization results.}
    \label{fig:visual}
\end{figure*}

%%%%%%%%%%% Tab: polyp_results %%%%%%%%%%%%%%%
\noindent \textbf{Qualitative Comparisons.}
Fig.~\ref{fig:visual} visualize the segmentation results of our method on the ERUS dataset.
In ultrasound videos, highly differentiated tumors exhibit diverse shapes and sizes, often with indistinct boundaries. 
Compared to other methods, our approach demonstrates superior robustness, enabling accurate localization and segmentation of tumors even in challenging scenarios. 
More importantly, our method offers higher detection rates and lower false detection rates, which are crucial for clinical decision-making.

%%%%%%%%%%%%%%%%%%%%%%%%%%% Fig 5 %%%%%%%%%%%%%%%%%%%%%%%%%
%%%%%%%%%%%%%%%%%%%%%%%%%%% Fig 5 %%%%%%%%%%%%%%%%%%%%%%%%%

\begin{figure}[t]
\centering
\begin{minipage}{\textwidth}
\centering
\begin{minipage}[t]{0.5\textwidth}
\makeatletter\def\@captype{table}
\centering
\caption{Ablation study on different components. \textcolor{red}{Red} denotes improvement, \textcolor{green}{green} denotes reduction.}
\renewcommand\tabcolsep{2pt}
\centering
\begin{tabular}{@{}lll|l@{}}
\toprule
Design    & MAE$\downarrow$ & Dice$\uparrow$ &FPS$\uparrow$                \\ \midrule
baseline  & 4.5  & 73.6 & \multirow{4}{*}{32.6}            \\ \cmidrule{1-3}
w/ CopyPaste       
          & 4.1$_{\color{red}{\downarrow0.4}}$  & 74.1$_{\color{red}{\uparrow0.5}}$ &             \\
w/ ArSDM   
          & 4.0$_{\color{red}{\downarrow0.5}}$  & 74.5$_{\color{red}{\uparrow0.9}}$ &             \\
w/ ASMA    & 3.2$_{\color{red}{\downarrow1.3}}$  & 75.6$_{\color{red}{\uparrow2.0}}$ &             \\ \midrule
w/ SCB    & 3.8$_{\color{red}{\downarrow0.7}}$  & 74.9$_{\color{red}{\uparrow1.3}}$ &             \\
\cmidrule{1-3}
w/ SCB+$\mathcal{L}_{aux}$ & 2.8$_{\color{red}{\downarrow1.7}}$ & 75.8$_{\color{red}{\uparrow2.2}}$ & 40.8$_{\color{red}{\uparrow8.2}}$\\ \cmidrule{1-3}
Ours      & \textbf{2.7}$_{\color{red}{\downarrow1.8}}$ & \textbf{77.6}$_{\color{red}{\uparrow4.0}}$ & 
\\ \bottomrule
\end{tabular}
\label{table:ablation_1}
\end{minipage}
\begin{minipage}[t]{0.47\textwidth}
\makeatletter\def\@captype{table}
\centering
\caption{Ablation study of ASMA as data augmentation.\\}
\renewcommand\tabcolsep{5pt}
\renewcommand\arraystretch{1.26}
\centering
\begin{tabular}{@{}llll@{}}
\toprule
Design   & MAE$\downarrow$ & Dice$\uparrow$ & Sen$\uparrow$                    \\ \midrule
SANet    & 4.5 & 73.2 & 72.0                   \\ 
w/ ASMA   & 3.6$_{\color{red}{\downarrow0.9}}$ & 74.6$_{\color{red}{\uparrow1.4}}$ & 74.8$_{\color{red}{\uparrow2.8}}$                   \\ \midrule
SLT-Net  & 3.0 & 75.2 & 74.0                   \\
w/ ASMA   & 2.7$_{\color{red}{\downarrow0.3}}$ & 77.0$_{\color{red}{\uparrow1.8}}$ & 77.7$_{\color{red}{\uparrow3.7}}$                   \\ \midrule
FLA-Net  & 3.2 & 74.8 & 78.0                   \\
w/ ASMA   & 2.9$_{\color{red}{\downarrow0.3}}$ & 76.4$_{\color{red}{\uparrow1.6}}$ & 77.2$_{\color{green}{\downarrow0.8}}$                   \\
\bottomrule
\end{tabular}
\label{table:ablation_2}
\end{minipage}
\end{minipage}
\end{figure}

\subsection{Ablation Study}
\textbf{Effectiveness of components.}
We investigate the contribution of each proposed component, as shown in Tab.~\ref{table:ablation_1}.
We employed a network without the ASMA and SCB modules as the "baseline". 
After individually integrating the designed ASMA and SCB into the network, significant performance improvements can be observed.
More importantly, the sparsing operations reduce the redundant computations, thereby accelerating the inference FPS.
Furthermore, combining these two modules results in a $1.8\%$ decrease in MAE, $4\%$ increase in Dice, and a $0.5\%$ increase in Sensitivity compared to the baseline. 
These results indicate that our method provides richer ultrasound features and mitigates noise interference during multi-frame context fusion. 
Additionally, we conducted a comparison with two advanced data augmentation techniques: CopyPaste\cite{ghiasi2021simple} and ArSDM\cite{du2023arsdm}. 
CopyPaste involves pasting foreground regions of each linear/convex-array mode onto a random convex/linear-array background, which yielded less performance improvement compared to our ASMA.
On the other hand, compared to ArSDM, which trains a diffusion model to generate simulation data, our method saves computational resources and achieves higher performance gains.

\noindent \textbf{Effectiveness of adaptive scanning mode augmentation.}
We further investigate the effectiveness of adaptive scanning mode augmentation by applying it to three different segmentation methods. 
As shown in Tab.~\ref{table:ablation_2}, all methods equipped with our proposed ASMA achieved observable segmentation performance without any extra cost during the inference stage.

\section{Conclusion}
\label{sec:conclusion}
In this paper, we explore the potential of computer-aided automated segmentation of colorectal cancer in endorectal ultrasound videos and contribute the first well-annotated endorectal ultrasound dataset with segmentation and infiltration depth staging labels.
Besides, we evaluate 10 state-of-the-art cancer segmentation methods on the proposed dataset and establish benchmark performance metrics.
Furthermore, we devise the first colorectal cancer segmentation model, ASTR, tailored for endorectal ultrasound videos and achieves state-of-the-art performance. 
We hope this work can inspire researchers and pave the way for future works on computer-aided automatic endorectal ultrasound diagnosis.

\section*{Acknowledgement}
This work was supported by NSFC with Grant No. 62293482, by the Basic Research Project No. HZQB-KCZYZ-2021067 of Hetao Shenzhen HK S\&T Cooperation Zone, by Shenzhen General Program No. JCYJ20220530143600001, by Shenzhen-Hong Kong Joint Funding No. SGDX20211123112401002, by the Shenzhen Outstanding Talents Training Fund 202002, by Guangdong Research Project No. 2017ZT07X152 and No. 2019CX01X104, by the Guangdong Provincial Key Laboratory of Future Networks of Intelligence (Grant No. 2022B121201\\0001), by the Guangdong Provincial Key Laboratory of Big Data Computing, \%The Chinese University of Hong Kong, Shenzhen CHUK-Shenzhen, by the NSFC 61931024\&12326610, by Key Area R\&D Program of Guangdong Province with grant No. 2018B030338001, by the Key Area R\&D Program of Guangdong Province with grant No. 2018B030338001, by the Shenzhen Key Laboratory of Big Data and Artificial Intelligence (Grant No. ZDSYS201707251409055), and by Tencent \& Huawei Open Fund.

%
% ---- Bibliography ----
%
% BibTeX users should specify bibliography style 'splncs04'.
% References will then be sorted and formatted in the correct style.
%
\bibliographystyle{splncs04}
\bibliography{main}

\section{Supplementary Material}
\begin{figure}[tbh]
    \centering
    \includegraphics[width=\linewidth]{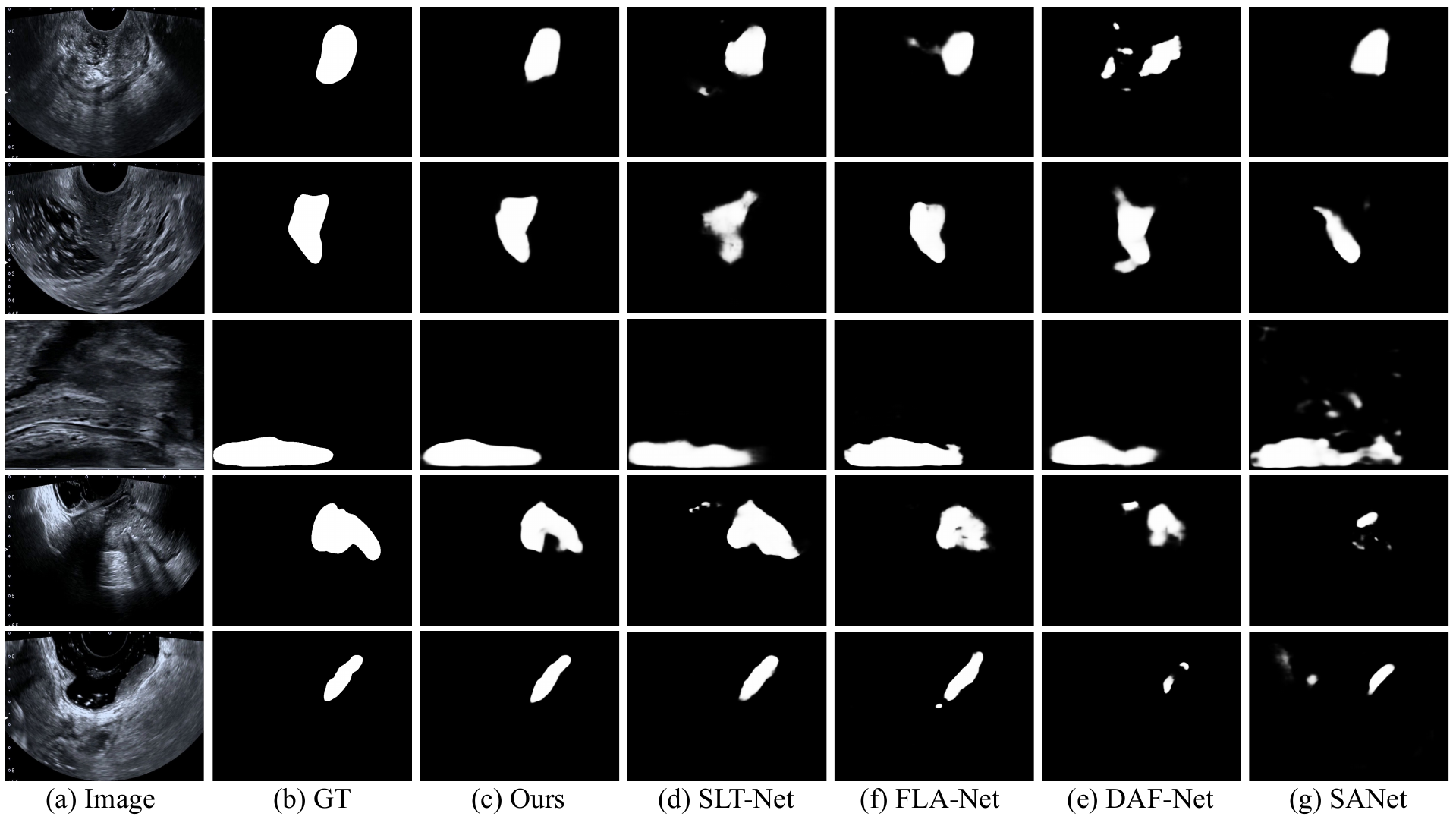}
    \caption{More samples of qualitative comparisons.}
    \label{fig:supp_visual}
\end{figure}

\begin{figure}[tbh]
    \centering
    \includegraphics[width=\linewidth]{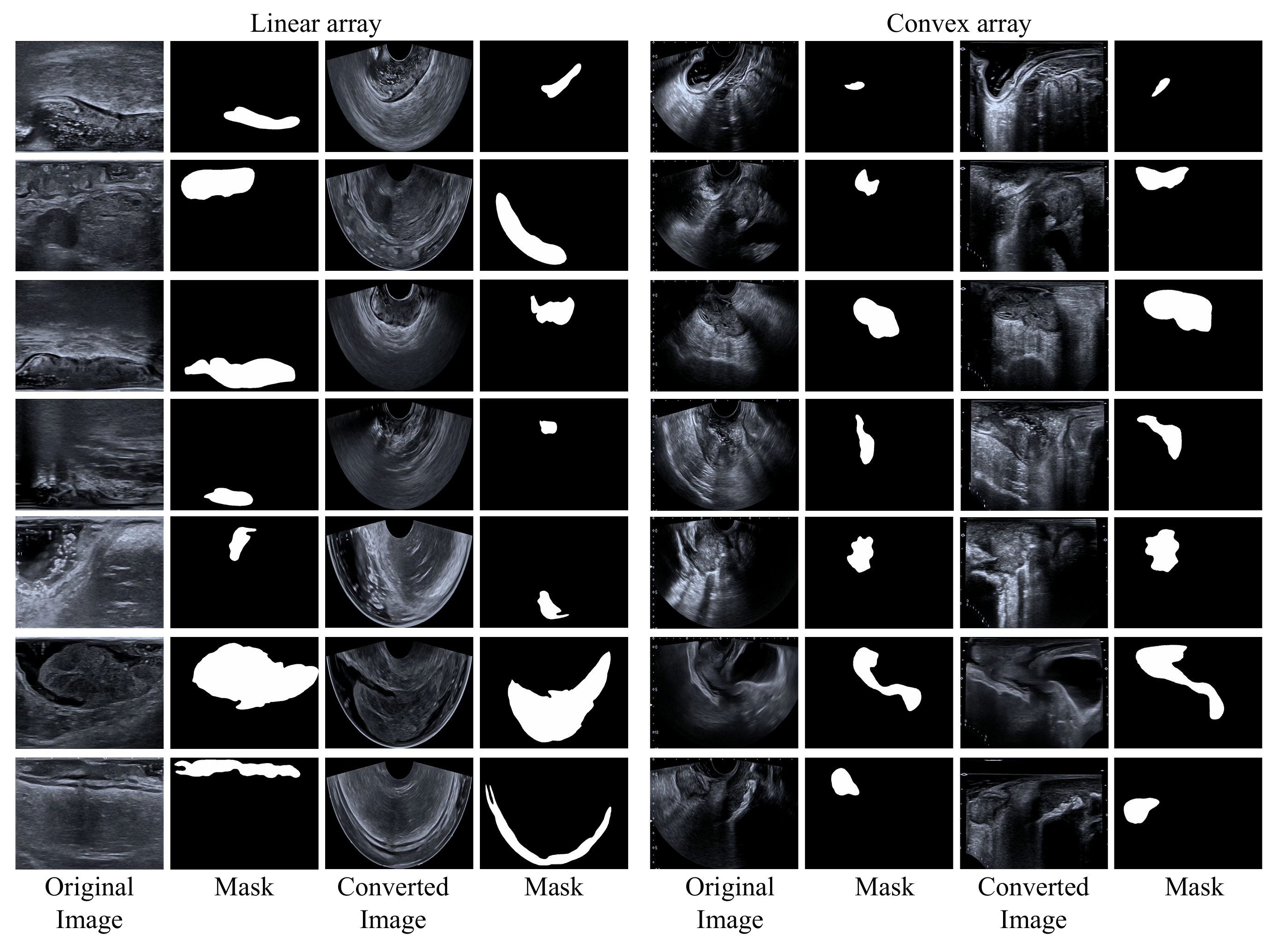}
    \caption{Sample images augmented by anatomic-aware transformation. images of linear-array/convex-array mode are transformed to images of convex-array/linear-array mode.}
    \label{fig:supp_asma}
\end{figure}

\end{document}